\documentclass[12pt]{article}
\usepackage{putex}

\makeatletter
\DeclareRobustCommand*{\bfseries}{%
  \not@math@alphabet\bfseries\mathbf
  \fontseries\bfdefault\selectfont
  \boldmath
}
\makeatother
\input{glyphtounicode}
\newcommand{\ndiv}{\nmid}
\def\extremum{\mathop{\rm extremum}\displaylimits}
\newcommand{\shove}[4][4in]{\setlength{\unitlength}{1in}\begin{picture}(0,0)(0,0)\put(#2,#3){\begin{minipage}{#1}\Black\raggedright #4\end{minipage}}\end{picture}\setlength{\unitlength}{1pt}}
\begin{document}

\preprint{PUPT-2523}

\title{A $p$-adic version of AdS/CFT}
\authors{Steven S. Gubser}
\institution{PU}{Joseph Henry Laboratories, Princeton University, Princeton, NJ 08544, USA}

\abstract{In this summary of my talk at Strings 2016, I explain how classical dynamics on an infinite tree graph can be dual to a conformal field theory defined over the $p$-adic numbers.  An informal introduction to $p$-adic numbers is followed by a presentation of results on holographic three- and four-point functions.  The simplicity of $p$-adic field theories and their similarity to ordinary field theories are illustrated through comparisons of holographic correlators and computations of simple loop diagrams on the field theory side.  I close with a discussion of challenges and directions for future work.}

\date{April 2017}
\maketitle

\section{Introduction}

The $p$-adic AdS/CFT correspondence \cite{Gubser:2016guj,Heydeman:2016ldy} puts a conformal field theory defined over the $p$-adic numbers in correspondence with a bulk theory defined on an infinite regular graph.  It has a number of notable precursors, including: the $p$-adic string \cite{Freund:1987kt}, where the worldsheet is replaced by just such a graph \cite{Zabrodin:1988ep} and the correlators giving rise to scattering amplitudes exhibit a $p$-adic version of conformal symmetry; studies in \cite{Manin:2002hn} of analogies between the BTZ black hole and quotients of the regular tree by a subgroup of the $p$-adic conformal group; and development in \cite{Harlow:2011az} of a class of stochastic processes on a regular tree, capturing aspects of eternal inflation where the far future is understood as a $p$-adic boundary.  My talk at Strings 2016 focused on the results of \cite{Gubser:2016guj}, in particular the classical action on the infinite regular graph, the appropriate definition of bulk-to-bulk and bulk-to-boundary propagators, three- and four-point functions, and the striking similarities between $p$-adic correlators and standard results on holographic correlators in ordinary AdS/CFT.  I also briefly discussed in my talk a remarkable non-renormalization property of kinetic terms in $p$-adic field theories, well known to practitioners (see for example \cite{Lerner:1989ty}) and related to the simplicity of the OPE expansion, in which descendant operators do not appear \cite{Melzer1989,Harlow:2011az}.

\section{$p$-adic numbers are naturally holographic}

Choose a prime number $p$.  Then the $p$-adic numbers, denoted $\mathbb{Q}_p$, are a completion of the rationals $\mathbb{Q}$ with respect to the so-called $p$-adic norm $|\cdot|_p$, defined so that
 \eqn{padicNorm}{
  \left| p^v {a \over b} \right|_p = p^{-v} \qquad\hbox{assuming}\qquad
   p \ndiv a \quad\hbox{and}\quad p \ndiv b \,,
 }
where $a$ and $b$ are integers with no common factors.  The key intuition is that the prime $p$ is small but not zero:
 \eqn{NormExamples}{
  |0|_p &= 0 \quad \hbox{(by fiat)}  \cr
  |a|_p &= 1 \quad \hbox{for $a=1,2,3,\ldots,p-1$}
    \cr
  |p|_p &= {1 \over p}\,, \qquad |p^2| = {1 \over p^2}\,, \qquad 
   |p^3| = {1 \over p^3} \quad\ldots
 }
A general non-zero $p$-adic number can be written as a formal series
 \eqn{padicSeries}{
  z = p^v \sum_{m=0}^\infty a_m p^m
 }
where each of the $p$-nary digits $a_m$ is chosen from the set $\{0,1,2,\ldots,p-1\}$ and the first one, $a_0$, is non-zero.  For example, if $p=2$, then
 \eqn{MinusOne}{
  -1 = {1 \over 1-2} = 1 + 2 + 2^2 + 2^3 + \ldots = \ldots 1111_2 \,,
 }
and the sum converges in the $2$-adic norm.  Saying that $\mathbb{Q}_p$ is the completion of $\mathbb{Q}$ with respect to $|\cdot|_p$ means that we are adjoining to $\mathbb{Q}$ an uncountable set of power series of the form \eno{padicSeries}, in lieu of the irrational numbers that we adjoin to $\mathbb{Q}$ to get the reals.  These power series are infinite when viewed as real numbers---but on the other hand, the ordinary irrationals are infinite with respect to any $p$-adic norm, because their $p$-nary expansions fail to terminate in the direction of negative powers of $p$.

A crucial property of $p$-adic numbers is ultrametricity:
 \eqn{Ultrametricity}{
  |x+y|_p \leq \sup\{|x|_p,|y|_p\} \,.
 }
The inequality \eno{Ultrametricity} implies the triangle inequality $|x+y|_p \leq |x|_p + |y|_p$ but is obviously stronger.  We will see that this ultrametric property implies a large simplification in holographic four-point functions, and it also is the key to the non-renormalization properties of kinetic terms in $p$-adic field theories.  Ultrametricity implies a failure of the so-called Archimedean property, which says that if $0 < |a| < |b|$, then $|b| < |na|$ for some integer $n$.  With the ultrametric property in hand, we have instead that if $|a|_p < |b|_p$, then $|na|_p < |b|_p$ for all $n$ because $|na|_p \leq |a|_p$.  Often, one sees constructions based on real numbers referred to as Archimedean and $p$-adic constructions as non-Archimedean.

Addition, multiplication, and multiplicative inverses can all be extended by continuity under $|\cdot|_p$ from the rationals to $\mathbb{Q}_p$, and one finds that $\mathbb{Q}_p$ is a field in the algebraic sense.  Perhaps surprisingly, the construction of $\mathbb{Q}_p$ fails if $p$ is composite.  Straightforward attempts lead not to a field, but to a ring which is not an integral domain, due to pairs of non-zero elements $x$ and $y$ with $xy=0$.

\begin{figure}
  \centerline{\includegraphics[width=4in]{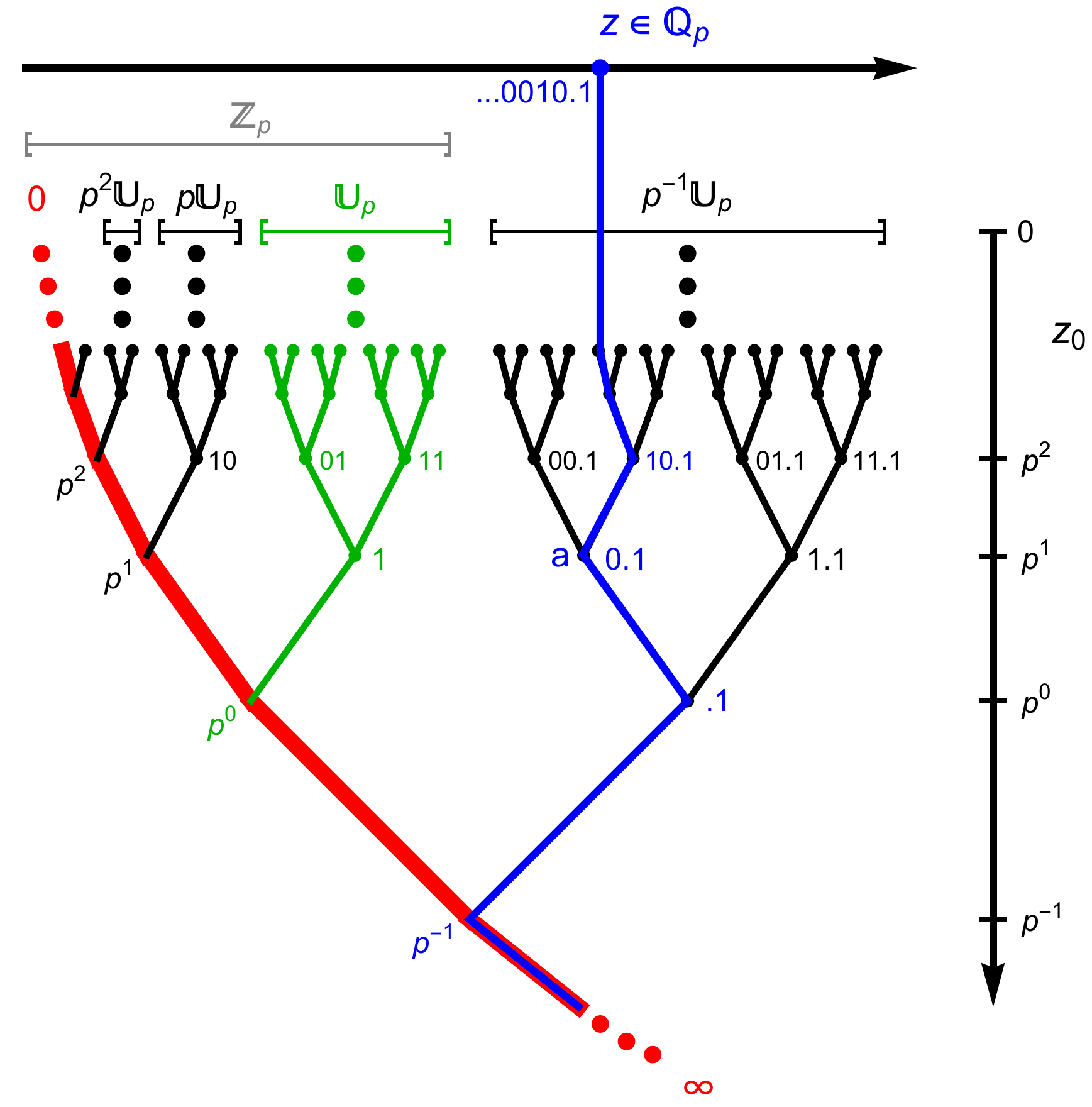}}
  \caption{The Bruhat-Tits tree with the trunk shown in red and a path shown in blue from $\infty$ to a non-zero $p$-adic number $z$.  Here we are setting $p=2$.  This and other figures are from \cite{Gubser:2016guj}.}\label{BruhatTits}
 \end{figure}

A first step toward $p$-adic AdS/CFT is the well-known Bruhat-Tits construction \cite{Bruhat:1972}, according to which $\mathbb{Q}_p$ is realized as the boundary of an infinite regular tree where each vertex has $p+1$ neighbors.\footnote{Actually, the boundary of the tree is the projective line $\mathbb{P}^1(\mathbb{Q}_p)$, which is $\mathbb{Q}_p$ together with a point at infinity.  This point at infinity plays a crucial role in our elementary account of the Bruhat-Tits tree.}  The standard description of this tree in the mathematical literature is given in terms of lattices in $\mathbb{Q}_p \times \mathbb{Q}_p$, but a more elementary account useful for our purposes can be based on the expansion \eno{padicSeries}.  Consider a tree like the one in figure~\ref{BruhatTits}.
We start by choosing arbitrarily an oriented path through the tree which we describe as the ``trunk.''  The path starts at a boundary point that we call $\infty$, and it ends at a boundary point that we call $0$.  All non-zero $p$-adic numbers can be naturally realized as the termination points of other oriented paths through the tree (with no back-tracking) starting from $\infty$.  It will be useful to introduce a depth coordinate $z_0$, similar to the depth coordinate in the Poincar\'e patch description of anti-de Sitter space, which takes values $p^v$ where $v \in \mathbb{Z}$.  This depth coordinate uniquely labels points along the trunk, and when we go off the trunk, every step we take is understood as ``upward'' in $z_0$.

Now let's trace how a chosen path upward through the tree corresponds to a $p$-adic number $z \in \mathbb{Q}_p$.  Our chosen path must depart from the trunk at some depth $z_0 = p^v$ (since otherwise we would just recover the path to the boundary point $0$), and we declare that $z$ will have $p$-adic norm $p^{-v}$.  Referring to \eno{padicSeries}, we see that the remaining information required to specify $z$ comprises the $p$-nary digits $a_m$, the first of which amounts to a $p-1$-fold chosen while the rest amount to $p$-fold choices.  Conveniently, our possible choices of path are specified by precisely this data.  When we first leave the trunk, we have $p-1$ different ways to do it (so only one way when $p=2$ as in figure~\ref{BruhatTits}).  At each subsequent vertex we must choose to continue up along one of $p$ possible branches.

Each node $a$ along our path can be understood as a rational approximation to $z$ obtained by truncating the $p$-adic expansion of $z$.  More precisely, it is useful to think of a node $a$ as specified by a pair $(z_0,z)$, where $z_0$ is its depth and $z$ is any $p$-adic number that can be reached by proceeding upward from that node.  Clearly, the choice of $z$ in the pair $(z_0,z)$ is highly non-unique: we must identify $(z_0,z) \sim (z_0,\tilde{z})$ precisely if $|z-\tilde{z}|_p \leq |z_0|_p$.  This is very different from Archimedean AdS/CFT \cite{Maldacena:1997re,Gubser:1998bc,Witten:1998qj}, where, for example, points in Euclidean ${\rm AdS}_2$ can be mapped bijectively to pairs $(z_0,z)$ where $z_0 \in \mathbb{R}^+$ and $z \in \mathbb{R}$.  But, as we will see, the formula for bulk-to-boundary propagators is naturally expressed in terms of $(z_0,z)$ in the $p$-adic context just as in the Archimedean context, and the ambiguity in the choice of $z$ will be just mild enough not to affect the formula for the propagator.

Although it is not apparent from our elementary description, the Bruhat-Tits tree can be realized as the quotient of the $p$-adic conformal group by its maximal compact subgroup.  This is well explained, for instance in \cite{brekke1988non}, so we will only present the claim.  The $p$-adic conformal group is the set of linear fractional transformations
 \eqn{LFT}{
  z \to {az + b \over cz + d} \,,
 }
where the matrix $M = \begin{pmatrix} a & b \\ c & d \end{pmatrix}$ has $p$-adic entries with $ac - bd \neq 0$.  These linear fractional transformations form the group ${\rm PGL}(2,\mathbb{Q}_p) = {\rm GL}(2,\mathbb{Q}_p) / \mathbb{Q}_p^\times$, where modding out by a non-zero $p$-adic number $\lambda \in \mathbb{Q}_p^\times$ corresponds to rescaling $M \to \lambda M$.  The maximal compact subgroup is ${\rm PGL}(2,\mathbb{Z}_p)$, where $\mathbb{Z}_p$ is the set of $p$-adic integers, corresponding to linear fractional transformations where the coefficients $a$, $b$, $c$, and $d$ are all in $\mathbb{Z}_p$.  Two equivalent characterizations of $\mathbb{Z}_p$ are:
 \begin{itemize}
  \item $\mathbb{Z}_p$ is the completion of the integers $\mathbb{Z}$ with respect to $|\cdot|_p$.
  \item $\mathbb{Z}_p$ is the subset of $\mathbb{Q}_p$ comprising elements $z \in \mathbb{Q}_p$ with $|z|_p \leq 1$.
 \end{itemize}
The second characterization shows that $\mathbb{Z}_p$ is analogous to the interval $[-1,1] \subset \mathbb{R}$.  This makes it at least somewhat intuitive that ${\rm PGL}(2,\mathbb{Z}_p)$ should be the maximal compact subgroup of ${\rm PGL}(2,\mathbb{Q}_p)$.  But because $\mathbb{Z}_p$ is uncountable, ${\rm PGL}(2,\mathbb{Z}_p)$ should be understood as three-dimensional just like ${\rm PGL}(2,\mathbb{Q}_p)$ itself.  So it makes sense that the quotient $T_p = {\rm PGL}(2,\mathbb{Q}_p)\slash {\rm PGL}(2,\mathbb{Z}_p)$ should be discrete.  This is in contrast to the Archimedean quotient ${\rm EAdS}_2 = {\rm SL}(2,\mathbb{R}) / {\rm U}(1)$, where the maximal compact subgroup $\mathbb{U}(1)$ is of lesser dimension than the conformal group ${\rm SL}(2,\mathbb{R})$.

\section{Correlators I: motivation and propagators}

The inception of the $p$-adic string in \cite{Freund:1987kt} hinged on the derivation of a $p$-adic variant of the Veneziano amplitude from a free-field four-point function integrated over the $p$-adic numbers.  Explicitly, the usual (crossing-symmetric) Veneziano amplitude can be expressed as
 \eqn{UsualVeneziano}{
  A^{(4)}_\infty &= \int_{\mathbb{R}} dz \, |z|^{k_1 \cdot k_2} |1-z|^{k_1 \cdot k_3}
   = \Gamma_\infty(-\alpha(s)) \Gamma_\infty(-\alpha(t)) \Gamma_\infty(-\alpha(u))
 }
where all $k_i^2 = 2$ (corresponding to tachyons), $s = -(k_1+k_2)^2$, $t = -(k_1+k_3)^2$, $u = -(k_1+k_4)^2$, $\alpha(X) = 1 + X/2$, and we define
 \eqn{Gammas}{
  \Gamma_\infty(\sigma) \equiv 2 \Gamma_{\rm Euler}(\sigma) \cos {\pi \sigma \over 2} \,.
 }
To define the $p$-adic string analog, we keep $k_i$, $s$, $t$, $u$, and $\alpha(X)$ all the same (in particular, all of them are real-valued, and we still insist that $k_i^2 = 2$) and set
 \eqn{padicVeneziano}{
  A^{(4)}_p &= \int_{\mathbb{Q}_p} dz \, |z|_p^{k_1 \cdot k_2} |1-z|_p^{k_1 \cdot k_3}
   = \Gamma_p(-\alpha(s)) \Gamma_p(-\alpha(t)) \Gamma_p(-\alpha(u))
 }
where
 \eqn{padicGamma}{
  \Gamma_p(\sigma) \equiv {1 - p^{\sigma-1} \over 1 - p^{-\sigma}} \,.
 }
The operation of $p$-adic integration can be defined so that
 \eqn{IntProp}{
  \int_{\mathbb{Z}_p} dz = 1 \qquad\hbox{and}\qquad
  \int_{\xi S} dz = |\xi|_p \int_S dz 
    \quad\hbox{for}\quad
    \hbox{$\xi \in \mathbb{Q}_p$, $S \subset \mathbb{Q}_p$}\,.
 }
Amazingly, $\prod_v \Gamma_v(z) = 1$ where the notation $\prod_v$ means to include $v=\infty$ and also $v=p$ for all primes.  So we get the ``adelic'' relation \cite{Freund:1987ck}
 \eqn{AdelicString}{
  \prod_v A^{(4)}_v = 1 \,.
 }
A much simpler adelic identity is
 \eqn{PrimeFactorization}{
  \prod_v |z|_v = 1 \qquad\hbox{for non-zero rational $z$}\,,
 }
which follows immediately from the prime factorization of $z$ as $z = s \prod_j p_j^{v_j}$ over some finite set of primes $p_j$, where $s = \pm 1$.  (We set $|z|_\infty = |z|$, the ordinary Archimedean absolute value.)

The integrand in \eno{padicVeneziano} can be understood essentially as the four-point function $\langle :\!e^{i k_1 X(z)}\!:\,:\!e^{i k_2 X(0)}\!:\,:\!e^{i k_3 X(1)}\!:\,:\!e^{i k_4 X(\infty)}\!: \rangle$ for a free field $X$.  The contribution of \cite{Zabrodin:1988ep} was to explain quite explicitly how one could start with a classical action for a scalar on the Bruhat-Tits tree and derive what was described there as a equivalent non-local effective theory on the boundary.  This account has a surprisingly modern feel to it and anticipates elements of the AdS/CFT correspondence.  In particular, the non-local effective theory would be understood in modern terms as a collection of power-law correlators characteristic of a $p$-adic conformal field theory---though a relatively trivial one since one can obtain all correlators of interest via Wick contraction starting from a two-point function.

In general, a $p$-adic conformal field theory can be understood to include some set of operators ${\cal O}(z)$, where $z \in \mathbb{Q}_p$ while ${\cal O}$ takes values in an algebra of operators on an ordinary Archimedean Hilbert space.  Two- and three-point functions of ${\cal O}(z)$ are completely fixed up to normalization by symmetry under the $p$-adic conformal group:
 \eqn{TwoThreePt}{
  \langle {\cal O}(z) {\cal O}(0) \rangle = {C_{\cal OO} \over |z|_p^{2\Delta}}\,, \qquad\qquad
  \langle {\cal O}(z_1) {\cal O}(z_2) {\cal O}(z_3) \rangle = 
    {C_{\cal OOO} \over |z_{12} z_{23} z_{31}|_p^\Delta} \,,
 }
where $z_{ij} \equiv z_i-z_j$ and $C_{\cal OO}$ and $C_{\cal OOO}$ are real numbers.  A first principles account of $p$-adic CFT can be found in \cite{Melzer1989}.

In place of
 \eqn{AdSAction}{
  S_{\rm bulk} = 
   \int_{{\rm EAdS}_2} d^2 z \, \sqrt{g} \left[ {1 \over 2} (\partial\phi)^2 + V(\phi) \right]
 }
it is natural to study
 \eqn{TreeAction}{
  S_{\rm tree} = \sum_{\langle ab \rangle} {1 \over 2} (\phi_a-\phi_b)^2 + 
   \sum_a V(\phi_a)
 }
where $\langle ab \rangle$ means that we are summing over adjacent vertices in the tree.  The linearized equation of motion involves only $m_p^2 = V''(0)$:
 \eqn{TreeLap}{
  (\square + m_p^2) \phi_a = 0 \qquad\hbox{where}\qquad 
   \square \phi_a = \sum_{\langle ab \rangle \atop \ a\ \rm fixed} (\phi_a - \phi_b) \,,
 }
and the bulk-to-bulk propagator, satisfying $(\square_a+m_p^2) G(a,b) = \delta_{a,b}$, is
 \eqn{GabAnsatz}{
  G(a,b) = {\zeta_p(2\Delta) \over p^\Delta} p^{-\Delta d(a,b)} \qquad\hbox{where}\qquad
   m_p^2 = -{1 \over \zeta_p(\Delta-1) \zeta_p(-\Delta)}
 }
and $d(a,b)$ is the number of steps from $a$ to $b$ on the tree.  We have defined the so-called local zeta function
 \eqn{zetap}{
  \zeta_p(\sigma) \equiv {1 \over 1 - p^{-\sigma}} \,.
 }
If we take a limit where $b = (y_0,y)$ goes to the boundary, and we rescale $G$ by a power of $y_0$ in the process, the result is the bulk-to-boundary propagator:
 \eqn{KDef}{
  K(z_0,z;y) = {\zeta_p(2\Delta) \over \zeta_p(2\Delta-1)} 
    {|z_0|_p^\Delta \over | (z_0,z-y) |_s^\Delta} \,,
 }
where $|(z_0,z-y)|_s = \sup\{ |z_0|_p,|z-y|_p \}$.  This is the formula we mentioned earlier in connection with the ambiguity of the parametrization $(z_0,z)$ of a bulk point.  We see that although $z-y$ is ambiguous, $|(z_0,z-y)|_s$ is not because $z$ is uncertain only by the addition of a quantity whose $p$-adic norm cannot exceed $|z_0|_p$.  So the bulk-to-boundary propagator is well defined by \eno{KDef}, and it has the standard property
 \eqn{NormalizationK}{
  \int_{\mathbb{Q}_p} dy \, K(z_0,z;y) = 1 \,.
 }

\begin{figure}
  \centerline{\includegraphics[width=3in]{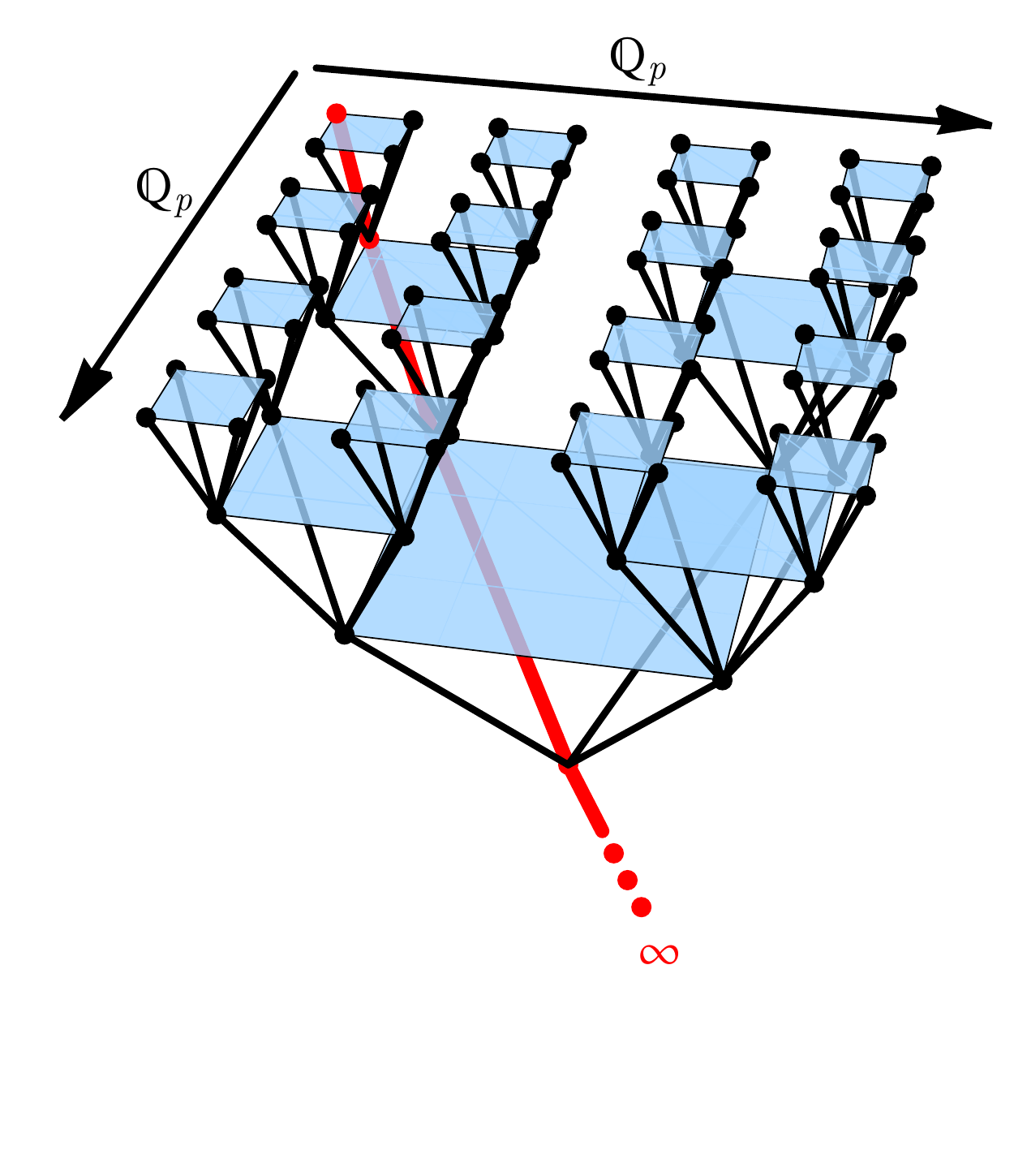}}
  \caption{The Bruhat-Tits tree for an unramified extension of $\mathbb{Q}_p$.  Here $p=2$ and the extension is quadratic.  The blue squares are present to guide the eye and to make clear the two-dimensional vector space structure of each layer of the tree.}\label{BruhatTits2}
 \end{figure}

Before turning to explicit calculations of holographic $n$-point functions, it pays to extend our field of view to include the so-called unramified extension $\mathbb{Q}_q$ of the $p$-adic numbers, where $q=p^n$ and $n$ is any integer greater than $1$.  $\mathbb{Q}_q$ is a field that contains $\mathbb{Q}_p$ and can be viewed as a vector space of dimension $n$ over $\mathbb{Q}_p$.  The field property of $\mathbb{Q}_q$ means in particular that given $\vec{x},\vec{y} \in \mathbb{Q}_q$ we have a commutative multiplicative operation that gives us the product $\vec{x}\vec{y} \in \mathbb{Q}_q$.  All this is analogous to the way we obtain the complex numbers from the reals: we adjoin $i = \sqrt{-1}$ to $\mathbb{R}$ to obtain a two-dimensional vector space over $\mathbb{R}$; but then complex multiplication gives us a rule for taking the product of any two elements of this vector space to give a new vector.  A non-trivial claim is that the $p$-adic norm extends uniquely to give a norm $|\vec{z}|_q$ on $\mathbb{Q}_q$ which still takes values $p^v$ where $v \in \mathbb{Z}$.  A fuller introduction to $\mathbb{Q}_q$ can be found in \cite{Gubser:2016guj}.  Suffice it to say here that a version $T_q$ of the Bruhat-Tits tree, where each vertex has $q+1$ nearest neighbors, can be constructed so that $\mathbb{Q}_q \cup \{\infty\}$ is its boundary, as illustrated in figure~\ref{BruhatTits2}.
Straightforward extension of the previous discussion shows that the bulk-to-boundary propagator on $T_q$ is
 \eqn{Kquad}{
  K(z_0,\vec{z};\vec{y}) = {\zeta_p(2\Delta) \over \zeta_p(2\Delta-n)}
    {|z_0|_p^\Delta \over |(z_0,\vec{z}-\vec{y})|_s^\Delta} \,,
 }
where now $|(z_0,\vec{z}-\vec{y})|_s = \sup\{ |z_0|_p,|\vec{z}-\vec{y}|_q \}$, and $\Delta$ is related to the mass $m_p^2$ by
 \eqn{padicMass}{
  m_p^2 = -{1 \over \zeta_p(\Delta-n) \zeta_p(-\Delta)} \,.
 }
By considering $T_q$ and its boundary $\mathbb{Q}_q \cup \{\infty\}$, we get closer to Archimedean ${\rm AdS}_{n+1}/{\rm CFT}_n$; indeed, we shall see quite a striking similarity in holographic Green's functions between the two constructions.  However, field theories on $\mathbb{Q}_q$ have similarities to low-dimensional Archimedean field theories no matter what $n$ is.  In particular, the natural notion of conformal group is still linear fractional transformations on $\mathbb{Q}_q$, now filling out the group ${\rm PGL}(2,\mathbb{Q}_q)$, in contrast to the Archimedean conformal group $SO(n+1,1)$ on $\mathbb{R}^n$.

An additional technical point is that it will often be useful to us to recast standard results on holographic correlators from Archimedean AdS/CFT in terms of one more zeta function, defined as
 \eqn{ZetaDef}{
  \zeta_\infty(\sigma) \equiv \pi^{-\sigma/2} \Gamma_{\rm Euler}(\sigma/2) \,.
 }
This function seems at first unmotivated, but it has two good properties: First, it allows us to write all local gamma functions as $\Gamma_v(\sigma) = \zeta_v(\sigma) / \zeta_v(1-\sigma)$, and second, when we express the Riemann zeta function as $\zeta_{\rm Riemann}(\sigma) = \prod_p \zeta_p(\sigma)$, we are naturally led to the modified function $\zeta_{\mathbb{A}}(\sigma) = \prod_v \zeta_v(\sigma) \equiv \zeta_\infty(\sigma) \zeta_{\rm Riemann}(\sigma)$, whose functional identity $\zeta_{\mathbb{A}}(\sigma) = \zeta_{\mathbb{A}}(1-\sigma)$ is a rephrasing of the identity $\prod_v \Gamma_v(\sigma) = 1$ that drives the adelic relation \eno{AdelicString} for $p$-adic string scattering amplitudes.

\section{Correlators II: three- and four-point functions}

\begin{figure}
  \centerline{\includegraphics[width=3in]{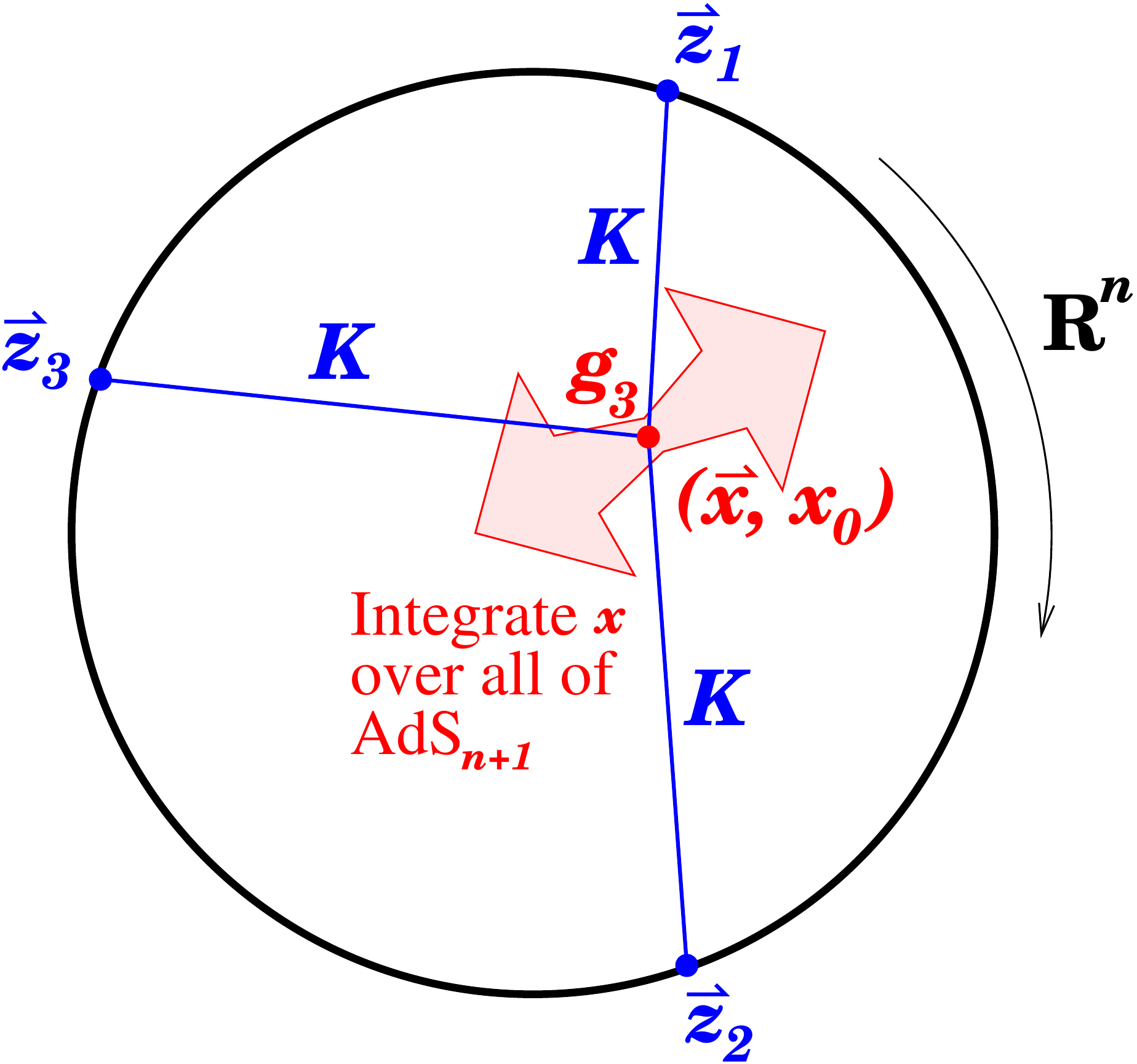}}
  \caption{The holographic calculation of the three-point function proceeds by integrating a cubic vertex over the whole bulk.}\label{ThreePtFunction}
 \end{figure}

First let's recall the standard story of correlation functions in Euclidean ${\rm AdS}_{n+1}/{\rm CFT}_n$ \cite{Gubser:1998bc,Witten:1998qj}: at tree level in the bulk,
 \eqn{GreenPrescription}{
  -\log \left\langle \exp\left\{
    \int_{\mathbb{R}^n} d^n z \, \phi_0(\vec{z}) {\cal O}(\vec{z}) 
    \right\} \right\rangle
   = \extremum_{\phi(z_0,\vec{z}) \to \phi_0(\vec{z})} S_{\rm bulk}[\phi] \,,
 }
where on the left hand side we have a coupling of $\phi_0$ to a singlet operator ${\cal O}$ in the field theory, and on the right hand side we have the bulk action subject to asymptotic boundary conditions on $\phi$, which are more precisely phrased as requiring
 \eqn{PhiAsymptotic}{
  \phi(z_0,\vec{z}) = z_0^{n-\Delta} \phi_0(\vec{z}) + \hbox{(sub-leading)} \qquad
   \hbox{for small $z_0$}\,.
 }
We have in mind the bulk action
 \eqn{BulkAction}{
  S_{\rm bulk} = \int_{{\rm EAdS}_{n+1}} d^{n+1} z \, \left[ {1 \over 2} (\partial\phi)^2
    + V(\phi) \right] \,.
 }
To get at the three-point function, we must differentiate \eno{GreenPrescription} by $\phi_0(\vec{z}_1)$, $\phi_0(\vec{z}_2)$, and $\phi_0(\vec{z}_3)$, and we set $g_3 = V'''(0)$.  Then, as illustrated in figure~\ref{ThreePtFunction}, the three-point function of ${\cal O}$ is
 \eqn{OOOreal}{
 \langle {\cal O}(\vec{z}_1) {\cal O}(\vec{z}_2) {\cal O}(\vec{z}_3) \rangle = 
   -g_3 \int_{{\rm AdS}_{n+1}} d^{n+1} x \, \sqrt{g} \, 
    K(x;\vec{z}_1) K(x;\vec{z}_2) K(x;\vec{z}_3) \,,
 }
where now the bulk-to-boundary propagator is
 \eqn{BBP}{
  K(x,\vec{z}) = {\zeta_\infty(2\Delta) \over \zeta_\infty(2\Delta-n)}
   {x_0^\Delta \over (x_0^2 + (\vec{z}-\vec{x})^2)^\Delta} \,.
 }
The first argument of $K$ in \eno{BBP} is a bulk point $x = (x_0,\vec{x})$.  The normalization of $K$ can be fixed through the requirement $\int_{\mathbb{R}^n} d^n z \, K(x;\vec{z}) = 1$.  Usually this normalization is expressed in terms of the Euler gamma function, but evidently it is simpler in terms of $\zeta_\infty$.  Conformal invariance dictates that
 \eqn{TPForm}{
  \langle {\cal O}(\vec{z}_1) {\cal O}(\vec{z}_2) {\cal O}(\vec{z}_3) \rangle = 
   {C_{\cal OOO}^{(\infty)} \over 
     |\vec{z}_{12}|^\Delta |\vec{z}_{23}|^\Delta |\vec{z}_{13}|^\Delta} \,,
 }
so the non-trivial part of the integral in \eno{OOOreal} amounts to finding the coefficient $C_{\cal OOO}^{(\infty)}$.  It is
 \eqn{COOO}{
  C_{\cal OOO}^{(\infty)} = -g_3 
     {\zeta_\infty(\Delta)^3 \zeta_\infty(3\Delta-n) \over 2\zeta_\infty(2\Delta-n)^3}
   = -g_3 \pi^{-n} {\Gamma_{\rm Euler}(\Delta/2)^3
    \Gamma_{\rm Euler}(3\Delta/2 - n/2) \over 2 \Gamma_{\rm Euler}(\Delta-n/2)} \,.
 }
In the last expression in \eno{COOO} we gave the standard form for $C_{\cal OOO}$ (cf.~for example \cite{Muck:1998rr,Freedman:1998tz}); note the explicit power of $\pi$, which conveniently disappears into the definition of $\zeta_\infty(\sigma)$.  This is one of many examples in which well-known formulas in AdS/CFT (and, indeed, in field theory generally) simplify when expressed in terms of $\zeta_\infty$ and related functions like $\Gamma_\infty$.

\begin{figure}
  \centerline{\includegraphics[width=3in]{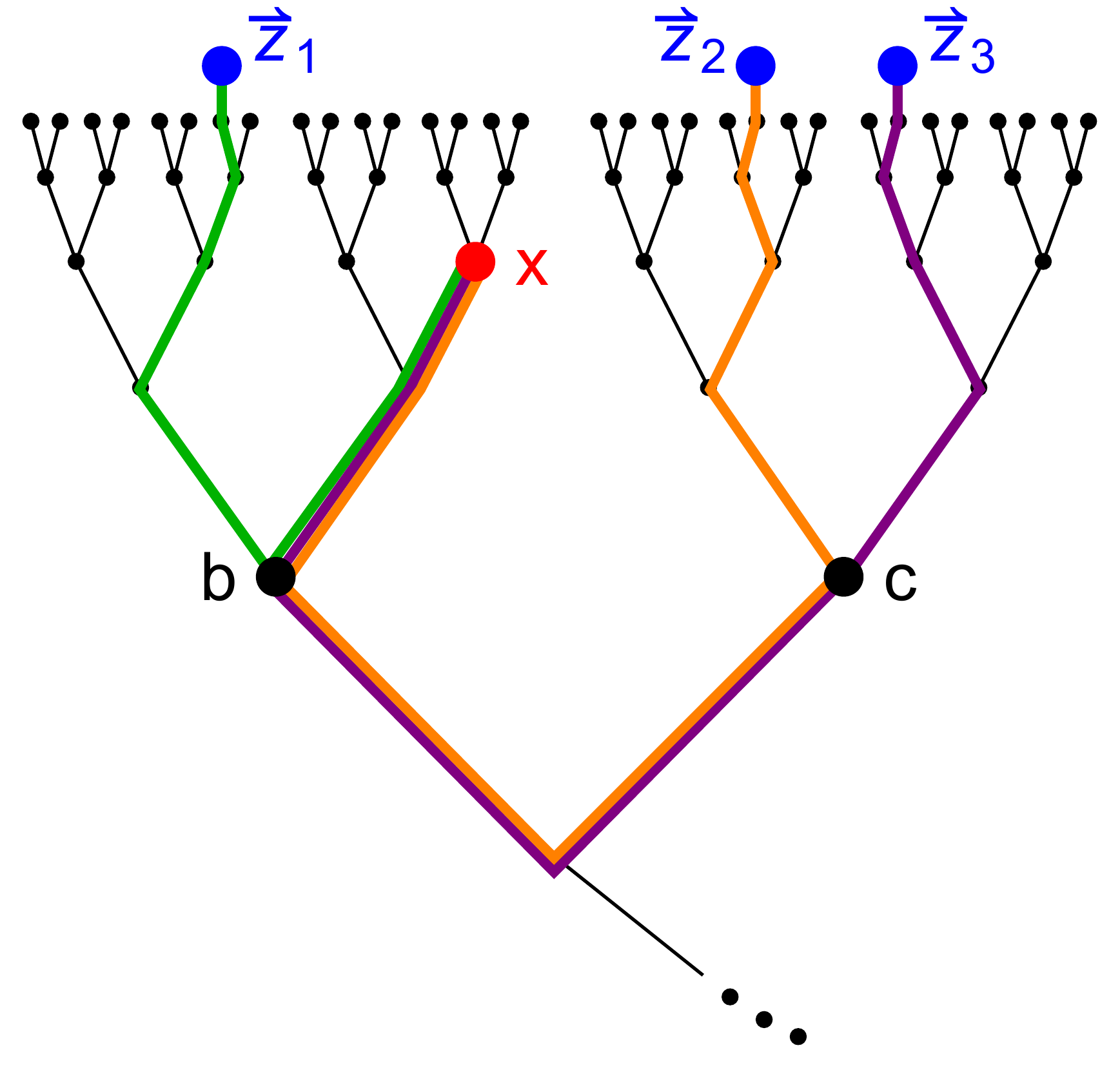}}
  \caption{The point $c$ is uniquely determined as the meeting point of paths from $\vec{z}_1$, $\vec{z}_2$, and $\vec{z}_3$.  The bulk point $x$ must be summed over the entire tree, and the point $b$ is the location where paths from $x$ to the other points diverge from the paths from all the $\vec{z}_i$ to $c$.}\label{ReSubway}
 \end{figure}

\begin{figure}
  \centerline{\includegraphics[width=3.2in]{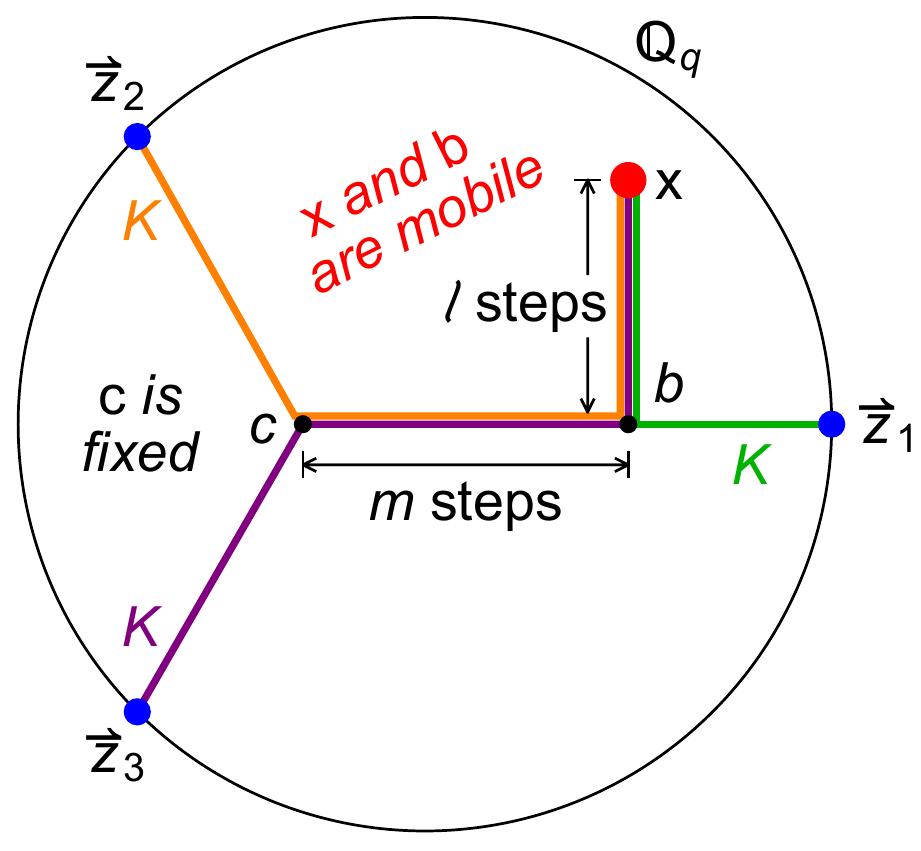}}
  \caption{A more abstract illustration of the sum of $x$ over the whole of $T_q$.  The variables $\ell$ and $m$ are the same as the ones appearing in \eno{ExplicitGfourth}.  I like to refer to this type of diagram as a subway diagram because of its resemblance to a map of a subway system.}\label{SubwayThree}
 \end{figure}

To do the corresponding calculation of a three-point correlator in $p$-adic AdS/CFT, we start with the action
 \eqn{TreeActionAgain}{
  S_{\rm tree} = \sum_{\langle ab \rangle} {1 \over 2} (\phi_a-\phi_b)^2 + 
   \sum_a V(\phi_a) \,,
 }
where now $a$ and $b$ run over all of $T_q$ and $\langle ab \rangle$ indicates an edge of $T_q$.  Define as before $g_3 = V'''(0)$, we find
 \eqn{OOOp}{
  \langle {\cal O}(\vec{z}_1) {\cal O}(\vec{z}_2) {\cal O}(\vec{z}_3) \rangle = 
    -g_3 \sum_x \prod_{i=1}^3 K(x,\vec{z}_i) = 
    {C_{\cal OOO}^{(p)} \over |\vec{z}_{12} \vec{z}_{23} \vec{z}_{13}|^\Delta} \,.
 }
The dependence on the $\vec{z}_i$ is fixed by $p$-adic conformal invariance, so as before the crux of the calculation is to find $C_{\cal OOO}^{(p)}$.  The final result is strikingly similar to the Archimedean case:
 \eqn{COOOp}{
  C_{\cal OOO}^{(p)} = -g_3 
     {\zeta_p(\Delta)^3 \zeta_p(3\Delta-n) \over \zeta_p(2\Delta-n)^3} \,.
 }
Two points are worthy of note before we enter into details of the computation:
 \begin{itemize}
  \item The absence of powers of $\pi$ is now inevitable: There is no place for them to come from, because the whole Green's function is based on a collection of variants of geometric series.
  \item We can almost anticipate the form of the answer by taking the Archimedean answer and replacing $\zeta_\infty$ by $\zeta_p$.  The only difference is a factor of $2$ in the denominator of \eno{COOO} which is not present in the denominator of \eno{COOOp}.  I believe this is related to the fact that the area of the unit sphere $S^{n-1}$ is $2/\zeta_\infty(n)$, whereas the volume of the set of units in $\mathbb{Q}_q$, namely the multiplicative group $\mathbb{U}_q$ of $p$-adic numbers with norm equal to $1$, is $1/\zeta_p(n)$.  But it's fair to say that a crisp understanding of this factor of $2$ is not presently within my grasp.
 \end{itemize}
In order to go from the infinite sum in \eno{OOOp} to the final result \eno{COOOp}, it helps to study the geometry of the propagators.  Paths from $\vec{z}_1$, $\vec{z}_2$, and $\vec{z}_3$ into the bulk meet (without back-tracking) at a uniquely determined bulk point $c$: See figure~\ref{ReSubway}.
We are instructed by \eno{OOOp} to run paths from the $\vec{z}_i$ not to $c$, but to some other point $x$ which is then summed over the entire tree.  So, except when $x=c$, there is in fact a great deal of back-tracking in the paths whose lengths enter into an overall power of $p$ that enters into the summand $\prod_{i=1}^3 K(x,\vec{z}_i)$.  A convenient strategy is to separate out all these back-tracking pieces by writing
 \eqn{SumSplit}{
  \sum_x \prod_{i=1}^3 K(x,\vec{z}_i) = \left[ \prod_{i=1}^3 K(c;\vec{z}_i) \right]
    \sum_x \hat{G}(c,b) \hat{G}(b,x)^3 \,,
 }
where we have introduced the un-normalized bulk-to-bulk propagator 
 \eqn{GhatDef}{
  \hat{G}(a,b) \equiv p^{-\Delta d(a,b)} \,.
 }
Conveniently, the factor in square brackets can be shown to be precisely $1/|\vec{z}_{12} \vec{z}_{23} \vec{z}_{13}|^\Delta$, so the normalization coefficient comes essentially from doing the sum over the back-tracking pieces, as illustrated in figure~\ref{SubwayThree}.
Explicitly,
 \eqn{ExplicitGfourth}{
  \sum_x \hat{G}(c,b) \hat{G}(b,x)^3 &= 3 \sum_{m=1}^\infty p^{-\Delta m} \left[ 1 + 
      \sum_{\ell=1}^\infty (q-1) q^{\ell-1} \left(p^{-\Delta\ell}\right)^3 \right]  \cr
   &\qquad{} +  
      \left[ 1 + \sum_{\ell=1}^\infty (q-2) q^{\ell-1}  \left(p^{-\Delta l}\right)^3 \right]
   = {\zeta_p(\Delta)^3 \zeta_p(3\Delta-n) \over \zeta_p(2\Delta)^3} \,.
 }

Having presented the calculation of the three-point correlator in detail, let me be much sketchier about the four-point function and refer the reader to \cite{Gubser:2016guj} for full details.  The dominant contribution to the four-point function in the limit $|\vec{z}_{12}||\vec{z}_{34}| \ll |\vec{z}_{13}||\vec{z}_{24}|$ takes the form
 \eqn{ContactDiagram}{
  \left\langle \prod_{i=1}^4 {\cal O}(\vec{z}_i) \right\rangle_{\rm leading \atop log} = 
    g_4 {\zeta(2\Delta)^4 \zeta(4\Delta-n) \over \zeta(2\Delta-n)^4 \zeta(4\Delta)}
    {\log(|\vec{z}_{12}||\vec{z}_{34}|/|\vec{z}_{13}||\vec{z}_{24}|) \over
     |\vec{z}_{13}|^{2\Delta} |\vec{z}_{24}|^{2\Delta}} \,.
 }
Here we may take $|\cdot| = |\cdot|_\infty$ or $|\cdot|_q$ depending if we want the result in Archimedean or $p$-adic ${\rm AdS}_{n+1}/{\rm CFT}_n$.  Likewise, we set $\zeta = \zeta_\infty$ or $\zeta_p$, and $\log = \log_e$ or $\log_p$ (the latter meaning the base $p$ logarithm, not the logarithm of a $p$-adic argument as in some of the literature).

For $v=\infty$ (ordinary AdS/CFT), \eno{ContactDiagram} is a restatement of results of \cite{DHoker:1999pj}; the full expression is available and is a complicated function of two independent cross-ratios,
 \eqn{CrossRatios}{
  u \equiv {|\vec{z}_{12}| |\vec{z}_{34}| \over |\vec{z}_{13}| |\vec{z}_{24}|} 
    \qquad\qquad
  \tilde{u} \equiv {|\vec{z}_{14}| |\vec{z}_{23}| \over |\vec{z}_{13}| |\vec{z}_{24}|}
    \,.
 }
For $v=p$ ($p$-adic AdS/CFT), the {\it full expression} (though still base just on the four-point contact diagram) for $u < 1$ is
 \eqn{FullFour}{
  \left\langle {\displaystyle\prod_{i=1}^4} {\cal O}(\vec{z}_i) \right\rangle &= 
   g_4 \left[ {1 \over \zeta_p(4\Delta)} \log_p u - 
     \left( {\zeta_p(2\Delta) \over \zeta_p(4\Delta)} + 1 \right)^2 + 3 \right]  \cr
    &\qquad\quad{} \times {\zeta_p(2\Delta)^4 \zeta_p(4\Delta-n) / \zeta_p(2\Delta-n)^4 \over
       |\vec{z}_{13} \vec{z}_{24}|^{2\Delta}}
 }
We would like to ask: Why is there no $\tilde{u}$ dependence in \eno{FullFour}, and where did $\log_p u$ come from?

The second question is easily answered: the key claim is that $u < 1$ implies $\tilde{u} = 1$ in a $p$-adic context.  We committed in \eno{ContactDiagram} and \eno{FullFour} to a setup where $u<1$, so the absence of any $\tilde{u}$ dependence is automatic.  To demonstrate that $\tilde{u} = 1$, we first note that if $\vec{U} = (\vec{z}_{12} \vec{z}_{34})/(\vec{z}_{13} \vec{z}_{24})$ and $\tilde{\vec{U}} = (\vec{z}_{14} \vec{z}_{23})/(\vec{z}_{13} \vec{z}_{24})$, then $\vec{U} + \tilde{\vec{U}} = 1$.  The inequality $u \equiv |\vec{U}|_p < 1$ now immediately implies $\tilde{u} \equiv |\tilde{\vec{U}}|_p = 1$ by the so-called tall isosceles property.  This property says that if $\vec{x} + \vec{y} + \vec{z} = 0$ in an ultrametric space, then $|\vec{x}| = |\vec{y}| \geq |\vec{z}|$, up to relabeling of $\vec{x}$, $\vec{y}$, and $\vec{z}$.  To prove the tall isosceles property, relabel so that $|\vec{z}| \leq |\vec{x}|$ and $|\vec{z}| \leq |\vec{y}|$.  Then $|\vec{x}| = |\vec{y} + \vec{z}| \leq \sup\{|\vec{y}|,|\vec{z}|\} = |\vec{y}|$.  Likewise $|\vec{y}| \leq |\vec{x}|$.  So $|\vec{x}| = |\vec{y}|$ as desired.

\begin{figure}
  \centerline{\includegraphics[width=3in]{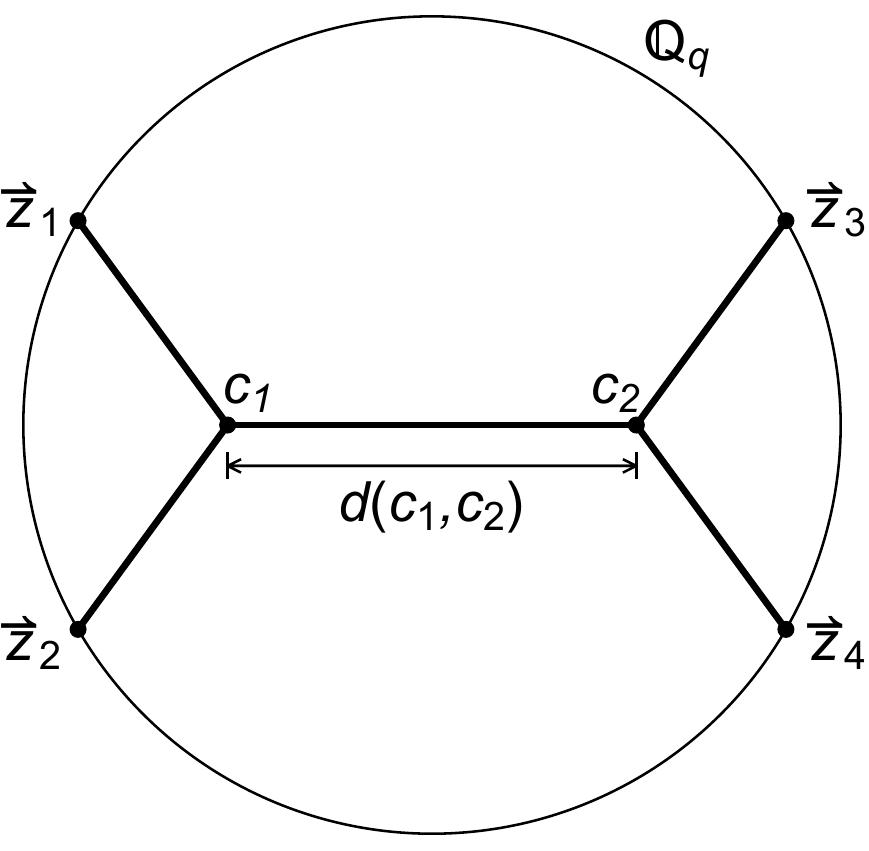}}
  \caption{Paths from the four points $\vec{z}_i$ into the bulk meet at points $c_1$ and $c_2$.}\label{CrossRatio}
 \end{figure}

\begin{figure}
  \centerline{\includegraphics[width=3in]{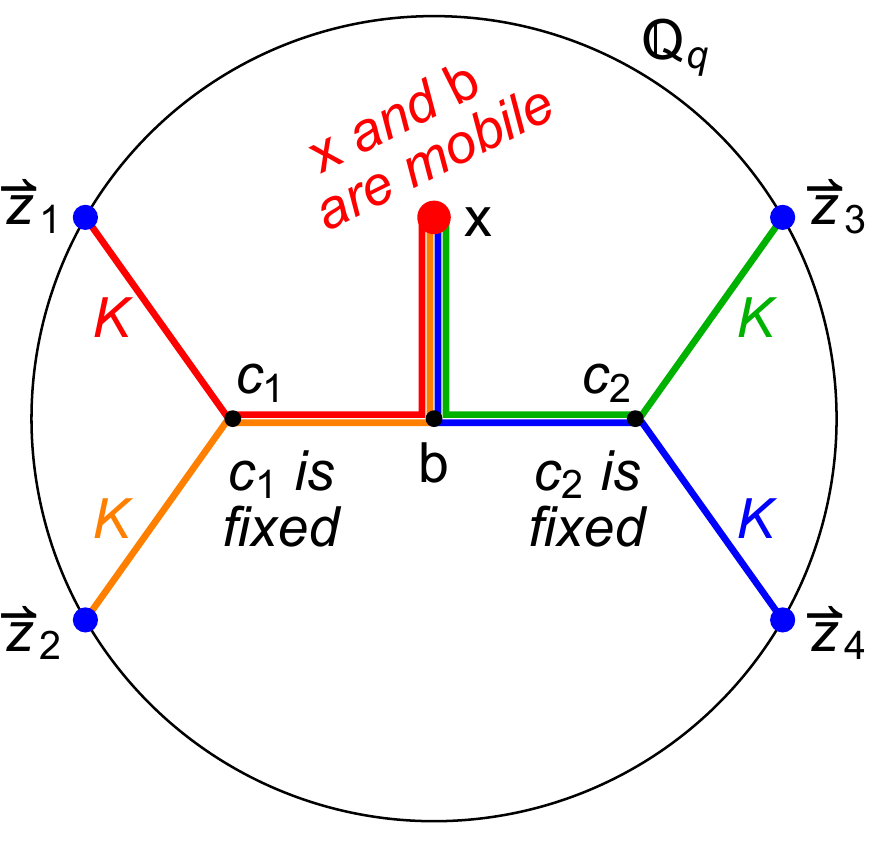}}
  \caption{The contributions to the holographic four-point function that give rise to logarithmic dependence on the cross-ratio $u$.}\label{FourMeetCenter}
 \end{figure}

\begin{figure}
  \centerline{\includegraphics[width=3in]{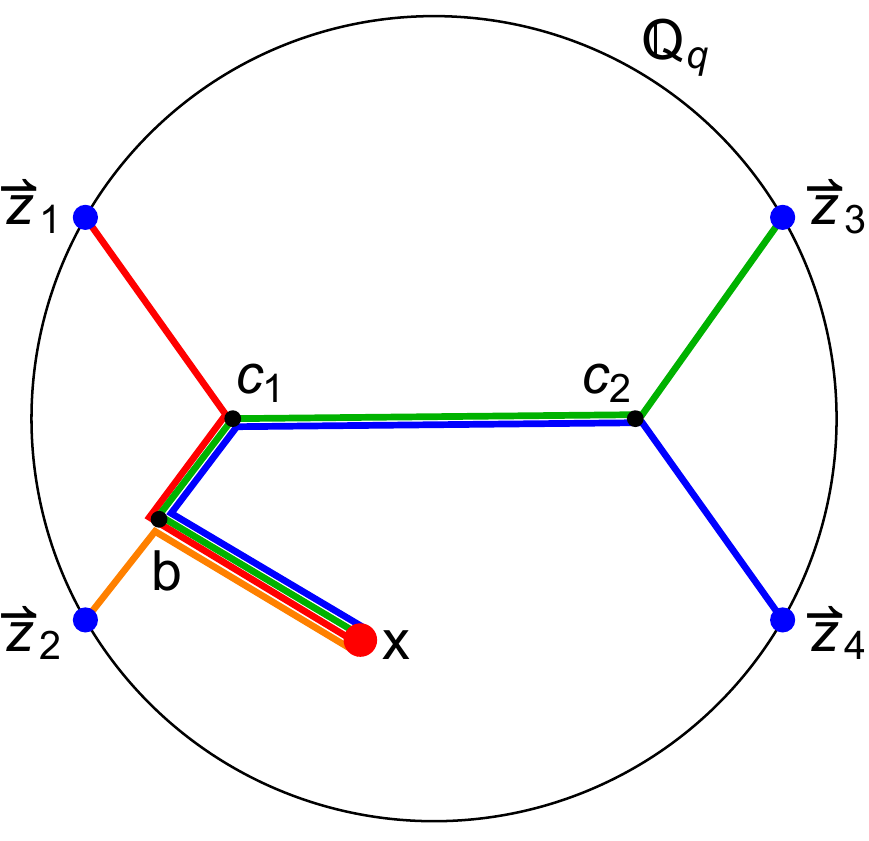}}
  \caption{Non-logarithmic contributions to the four-point function.}\label{FourMeetLeg}
 \end{figure}

To understand the factor of $\log_p u$, we need to once again study the geometry of paths on the tree $T_q$.  In the limit we are most interested in, with $|\vec{z}_{12}||\vec{z}_{34}| \ll |\vec{z}_{13}||\vec{z}_{24}|$, paths from $\vec{z}_1$ and $\vec{z}_2$ meet in the bulk at a point $c_1$ which is connected by a path of many steps to the point $c_2$ where one must split off in different directions to get to $\vec{z}_3$ or $\vec{z}_4$.  The length of the path from $c_1$ to $c_2$ is $d(c_1,c_2) = -\log_p u$: see figure~\ref{CrossRatio}.
Now, when we calculate the four-point function as
 \eqn{OOOOp}{
  \left\langle \prod_{i=1}^4 {\cal O}(\vec{z}_i) \right\rangle = -g_4
    \sum_x \prod_{i=1}^4 K(x,\vec{z}_i) \,,
 }
one class of contributions to the right hand side comes from points $x$ on branches that emerge from the path between $c_1$ and $c_2$, as shown in figure~\ref{FourMeetCenter}.
The number of such branches is essentially $d(c_1,c_2)$, and they are all equivalent, and so together they give rise to a term multiplied by $-\log_p u$ in the holographic four-point function \eno{FullFour}.  Additional non-logarithmic terms arise from contributions where $x$ is on a branch that emerges from an external leg, as shown in figure~\ref{FourMeetLeg}.

Note that, appearances to the contrary from the figures, we are considering {\it only} the contact diagram contribution to the four-point function.  If $g_3 \neq 0$, there are also in principle contributions proportional to $g_3^2$ from exchange diagrams.\footnote{Exchange diagram contributions have recently been studied in \cite{Gubser:2017tsi}.}  Logarithmic behavior suggests a small anomalous dimension for the operator ${\cal O}^2$, according to the same logic that comes up in ordinary AdS/CFT.  Here is a brief summary:
 \begin{itemize}
  \item Suppose $\Delta_{{\cal O}^2} = 2\Delta_{\cal O} + \delta$ with $\delta$ small.
  \item In an OPE expansion, ${\cal O}(\vec{z}_1) {\cal O}(\vec{z}_2) \sim |\vec{z}_{12}|^{-\delta} {\cal O}^2(\vec{z}_1)$ if $|\vec{z}_{12}|$ is small.
  \item Using the OPE expansion twice in ways justified by our limit $|\vec{z}_{12}||\vec{z}_{34}| \ll |\vec{z}_{13}||\vec{z}_{24}|$, we have
 \eqn{FourLog}{
  \left\langle \prod_{i=1}^4 {\cal O}(\vec{z}_i) \right\rangle \sim |\vec{z}_{12} \vec{z}_{34}|^\delta \langle {\cal O}^2(\vec{z}_1) {\cal O}^2(\vec{z}_3) \rangle \sim {|\vec{z}_{12} \vec{z}_{34}|^\delta \over |\vec{z}_{13}|^{4\Delta+2\delta}} \approx {1 - \delta \log u \over |\vec{z}_{13} \vec{z}_{24}|^{2\Delta}}
 }
 \end{itemize}

\section{An excursion into $p$-adic field theory}

For the most part I have focused on the holographic calculations of Green's functions that form the foundation of AdS/CFT in a $p$-adic context.  The story would be incomplete however without some account of $p$-adic field theory.  In point of fact, $p$-adic AdS/CFT {\it is} quite an incomplete subject, because it lacks (even as of time of writing) a clear and complete account of a dual pair of a bulk action and a corresponding boundary field theory.  Elements of perturbative $p$-adic field theory can be understood straightforwardly, including the non-renormalization of kinetic terms mentioned previously.  The current section barely scratches the surface of the large topic of $p$-adic field theory.  The recent work \cite{Gubser:2017vgc} provides a more complete introduction as well as a list of some of the standard references.  The points discussed in this section are not new; for the most part they may be found in \cite{Lerner:1989ty}.

To get started with field theory, we need some notion of a Fourier transform over $\mathbb{Q}_q$, which is as before the unramified $n$-dimensional extension of $\mathbb{Q}_p$, standing approximately in lieu of $\mathbb{R}^n$.  We decompose a complex-valued function $f$ over $\mathbb{Q}_q$ into Fourier modes according to
 \eqn{Fourier}{
  f(x) = \int_{\mathbb{Q}_q} d^n k \, \chi(kx) \hat{f}(k) \qquad\hbox{where}\qquad
   \chi(\xi) = e^{2\pi i [\xi]} \,,
 }
and $[\xi]$ is a rational number which intuitively is the fractional part of $\xi$.  The details of the construction of $\chi$ are slightly tricky.  Let's be satisfied with the statement that $\chi$ maps $\mathbb{Q}_q^\times$ (the non-zero $p$-adic numbers) to $S^1 \subset \mathbb{C}$ (the complex numbers with unit modulus) in such a way that $\chi(\xi_1 + \xi_2) = \chi(\xi_1) \chi(\xi_2)$.  The mathematical term for $\chi$ is an additive character, while physicists might visualize $\chi$ as a plane wave over $\mathbb{Q}_q$.

Suppose we have a two-point function for a field $\phi$ in a $p$-adic field theory which takes the form
 \eqn{Gpp}{
  G_{\phi\phi}(k) = \langle \phi(k) \phi(-k) \rangle = {1 \over |k|_q^s + r} \,.
 }
We require $r$ and $s$ to be real, whereas $k \in \mathbb{Q}_q$, and $\phi$ is realized as an operator in an Archimedean Hilbert space.  If we add quartic interactions for $\phi$, say with an action
 \eqn{SphiFour}{
  S = \int_{\mathbb{Q}_q} d^n k \, 
   {1 \over 2} \phi(-k) (|k|_q^s + r) \phi(k) + 
    \int_{\mathbb{Q}_q} d^n x \, {\lambda \over 4!} \phi(x)^4 \,,
 }
then we encounter divergences whose forms are familiar from our experience of Archimedean field theories: for example, imposing a hard cutoff, the simple loop diagram pictured has amplitude proportional to
 \eqn{LambdaDivergence}{
  \shove{-0.5}{0}{\includegraphics[width=0.4in]{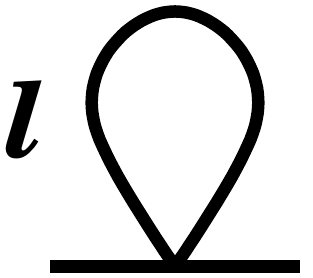}}\qquad \int_{\mathbb{Q}_q}^\Lambda {d^n \ell \over |\ell|_q^s + r} \sim \Lambda^{n-s} \,,
 }
where $\int_{\mathbb{Q}_q}^\Lambda$ means that we integrate over all $\ell \in \mathbb{Q}_q$ with $|\ell|_q < \Lambda$.  So for instance if $n=4$ and $s=2$, we get a quadratic divergence just like we would expect in $\mathbb{R}^4$.  A key difference is that the parameter $s$ is freely adjustable in $p$-adic field theories.  This is connected to a non-renormalization property: all loop corrections act only on the non-derivative part of the action, so that the effective action we would get starting from \eno{SphiFour} has the form
 \eqn{SeffPadic}{
  S_{\rm eff} = 
   \int d^n k \, {1 \over 2} \phi(-k) |k|_q^s \phi(k) +
   \int d^n x \, V_{\rm eff}(\phi) \,.
 }

\begin{figure}\hskip3in{\hbox{
  \shove{-1.5}{1}{\includegraphics[width=1.3in]{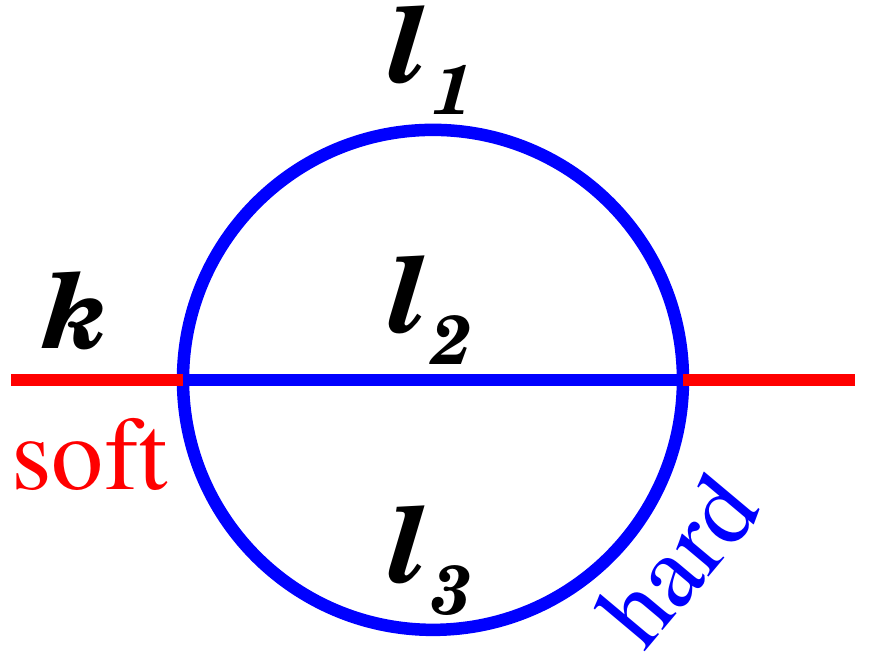}}
  \includegraphics[width=2in]{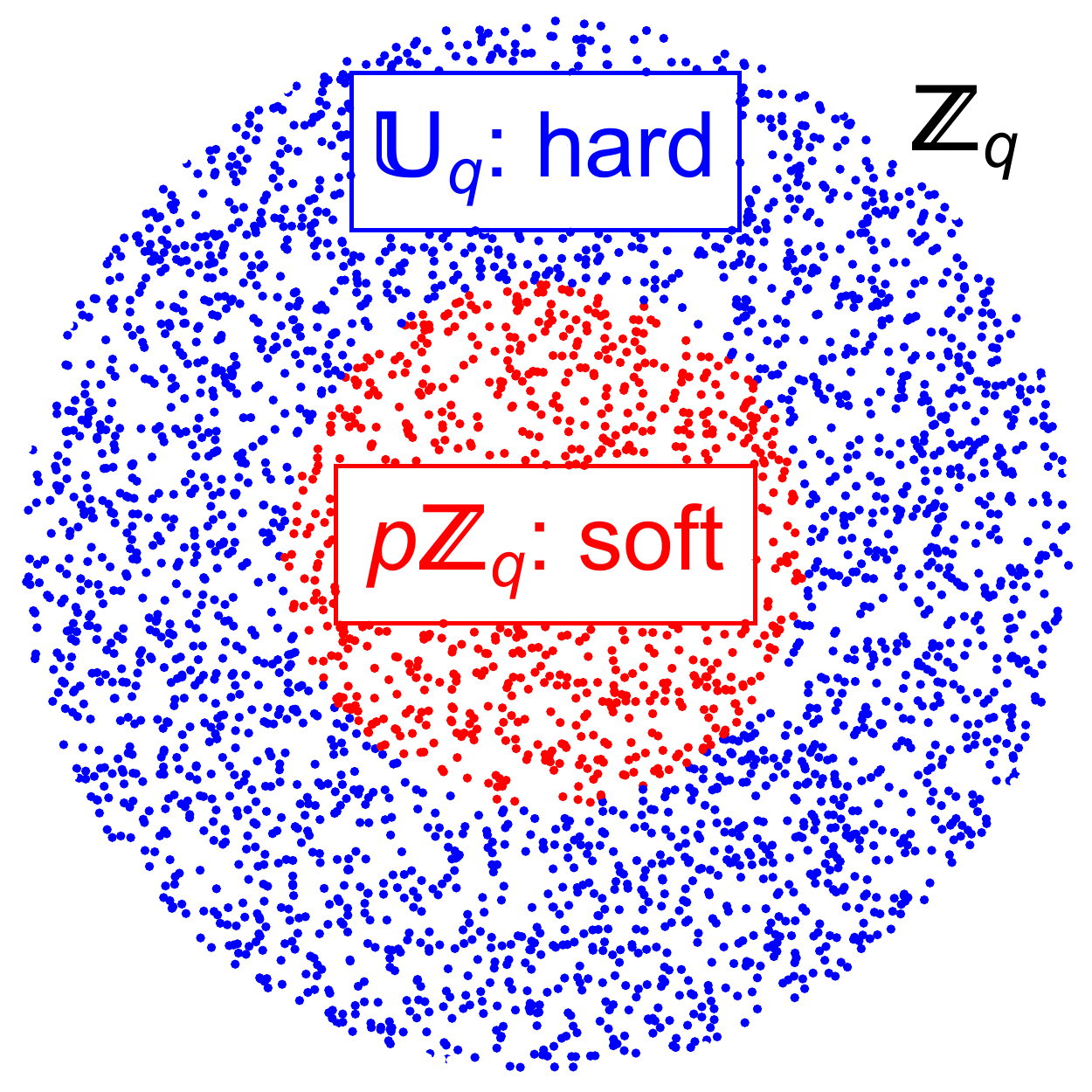}}}
  \caption{Left: the ``underground'' diagram, which in ordinary $\phi^4$ theory gives the leading correction to the kinetic term, has an entirely momentum-independent amplitude in $p$-adic field theory.  Right: A graphical depiction of integrating out a momentum shell.  Blue dots are elements of $\mathbb{U}_q$, and red dots are elements of $p\mathbb{Z}_q$.}\label{MomentumShell}
 \end{figure}

As an elementary example of the non-renormalization property of kinetic terms, consider in a Wilsonian framework the ``underground'' diagram which in Archimedean $\phi^4$ theory would give the leading correction to the kinetic term.  A Wilsonian development is particularly natural in $p$-adic field theory because $\mathbb{Q}_q$ decomposes into a disjoint union of concentric shells:
 \eqn{Qunion}{
  \mathbb{Q}_q = \bigsqcup_{v \in \mathbb{Z}} p^v \mathbb{U}_q \,,
 }
where as above $\mathbb{U}_q$ is the multiplicative group of elements of $\mathbb{Q}_q$ with unit norm.  The Wilsonian coarse-graining step consists in momentum space of starting with a cutoff $|\ell|_q \leq 1$ on all momenta, and integrating out momenta with $|\ell|_q=1$ to find an effective theory for external leg momenta satisfying $|k|_q \leq 1/p$.  Figure~\ref{MomentumShell} shows how we assign ``hard'' momenta $\ell_i$ with $|\ell_i|_q=1$ to internal propagators and a ``soft'' momentum $k$ with $|k|_q \leq 1/p$ to the external legs.
The amplitude for the underground diagram is proportional to
 \eqn{Itwo}{
  I_2(k) \equiv \int_{\mathbb{U}_q} 
    {d^n\ell_1 \, d^n\ell_2 \, d^n\ell_3 \over (1+r)^3}
    \delta(\textstyle{\sum_{i=1}^3 \ell_i} - k) = I_2(0) \,.
 }
We enforce the constraints $|\ell_i|_q = 1$ by integrating each of the $\ell_i$ over $\mathbb{U}_q$.  Hence the denominator, which ordinarily would be $\prod_{i=1}^3 (|\ell_i|_q^s + r)$, simplifies to $(1+r)^3$.  The delta function in the numerator enforces momentum conservation at both vertices.  To see that the whole integral is independent of $k$, we can make the $u$-substitution $\tilde\ell_3 = \ell_3 - k$.  Because $|k|_q < 1$, we have by the tall isosceles property $|\ell_3|_q = 1$ iff $|\tilde\ell_3|_q = 1$.  Thus $\ell_3 \to \tilde\ell_3$ is a bijection from $\mathbb{U}_q$ to itself.  The effect of the $u$-substitution is the same as if we had set $k=0$, and the $k$-independence of $I_2$ follows immediately.  This argument is readily generalized to an arbitrary perturbative amplitude \cite{Lerner:1989ty}, and the non-renormalization of kinetic terms follows immediately.  One way to understand this non-renormalization property is that non-derivative operators $\phi^m$ mix among themselves but not with $\phi \partial^\ell \phi$ like in the real case.  Thus, I suspect, it is closely related to the simplicity of the four-point function: The OPE ${\cal O}(\vec{z}) {\cal O}(0)$ is presumably very sparse, with no descendants.  The sparseness of the OPE is quite evident from the holographic four-point function \eno{FullFour}.  See also in this connection the general arguments of \cite{Melzer1989} as well as the stochastic construction of \cite{Harlow:2011az}.

\section{Conclusions and future directions}

At a technical level, $p$-adic AdS/CFT feels like a development that was waiting to happen.  It is quite natural to employ an extra dimension in describing $p$-adic numbers, similar to the depth direction $z_0$ in anti-de Sitter space.  In a $p$-adic context, $z_0$ is the $p$-adic accuracy of a truncated $p$-nary expansion of a boundary point $z \in \mathbb{Q}_p$.  What takes getting used to is that the bulk geometry is a discrete graph, the Bruhat-Tits tree.  So instead of a two-derivative action in the bulk, the simplest construction involves nearest neighbor interactions.  There is no sense in which we need to take a continuum limit of the discrete action on the graph: $p$-adic conformal symmetry is fully present as an isometry group of the graph, and it fixes the functional form of two-point correlators and three-point correlators in the boundary theory just as in ordinary conformal field theory.

Given how different the bulk theories are between the tree graph used in $p$-adic AdS/CFT and the smooth anti-de Sitter geometry of ordinary AdS/CFT, it is quite striking that the normalization of correlation functions is so similar between the two cases.  We gave details mainly for the three-point function, but we also explained how the leading logarithmic contribution to the four-point function is essentially the same in the $p$-adic and Archimedean versions of AdS/CFT.  Two-point functions have some additional subtleties due to divergent local terms, and the interested reader may consult \cite{Gubser:2016guj} for a full account.  In broad terms (and with significant caveats), the recipe for converting standard holographic correlators into $p$-adic ones is to write the former in terms of the local zeta function $\zeta_\infty$ and then replace $\zeta_\infty \to \zeta_p$ while also replacing ordinary absolute values with $p$-adic norms.  Such a simple recipe hints at some adelic simplifications extending the results of \cite{Freund:1987ck}, but at time of writing the adelic story is very incompletely understood.

Some obvious challenges going forward are to give an account of fluctuating geometry in the tree graph (and progress in this direction has been reported in \cite{Gubser:2016htz}), to extend the $p$-adic story from Euclidean to Lorentzian signature, and to find explicit $p$-adic AdS/CFT dual pairs.

As the twenty-first century advances without even a hint of supersymmetry, we are entitled to ask whether some really different ideas about the geometric foundations of physics might be required from the ones string theorists usually pursue.  We can certainly see from $p$-adic AdS/CFT as it stands now that ultrametric number systems, discrete geometries, and the holographic principle mesh very naturally.  Can we go further and envision a transition between smooth Archimedean geometry and ultrametric or discrete geometry at a scale comparable to the Planck length?  Or could there instead be an overarching geometry like superspace underlying fundamental physics, but with the additional data beyond ordinary Archimedean dimensions being ultrametric rather than fermionic in nature?  Similar aspirational ideas have been advanced by $p$-adic proponents in the past; see for example \cite{Manin13} or the introduction to \cite{vladimirov1994p}.  The challenge is to use ultrametric number systems to break into the tightly defined structure of local quantum field theory in a way that will go beyond formal results and help inform our understanding of fundamental physics.  Tokens of progress toward this goal include formulations of theories which are valid irrespective of the underlying number system.  The similarities between $p$-adic and Archimedean holographic correlators are encouraging hints of the existence of such a formulation of AdS/CFT.

\section*{Acknowledgments}

I am grateful to the organizers of Strings 2016 for having put together a stimulating conference and for hosting me as a speaker.  I particularly thank J.~Knaute, S.~Parikh, A.~Samberg, and P.~Witaszczyk for a stimulating collaboration.  This work was supported in part by the Department of Energy under Grant No.~DE-FG02-91ER40671.

\bibliographystyle{ssg}
\bibliography{padic}
\end{document}